\documentclass[twocolumn,showpacs,showkeys,amsmath,amssymb]{revtex4}

\usepackage{graphicx}

\begin{document}

\title{Alternatives to Schwarzschild in the weak field limit of General Relativity}

\author{V. Bozza$^{a,b}$, A. Postiglione$^{c,d}$}

\affiliation{$^a$ Dipartimento di Fisica ``E.R. Caianiello'',
Universit\`a di Salerno, Via Giovanni Paolo II 132,
I-84084 Fisciano (SA), Italy.\\$^b$ Istituto Nazionale di Fisica Nucleare, Sezione di Napoli, Italy.\\$^c$ Dipartimento di Fisica ``E. Amaldi'',Universit\`a di Roma Tre, Via della Vasca Navale 84, 00149 Roma, Italy.\\$^d$ Istituto Nazionale di Fisica Nucleare, Sezione di Roma Tre, Italy.}

\date{\today}
\begin{abstract}
The metric outside an isolated object made up of ordinary matter is bound to be the classical Schwarzschild vacuum solution of General Relativity. Nevertheless, some solutions are known (e.g. Morris-Thorne wormholes) that do not match Schwarzschild asymptotically. On a phenomenological point of view, gravitational lensing in metrics falling as $1/r^q$ has recently attracted great interest. In this work, we explore the conditions on the source matter for constructing static spherically symmetric metrics exhibiting an arbitrary power-law as Newtonian limit. For such space-times we also derive the expressions of gravitational redshift and force on probe masses, which, together with light deflection, can be used in astrophysical searches of non-Schwarzschild objects made up of exotic matter. Interestingly, we prove that even a minimally coupled scalar field with a power-law potential can support non-Schwarzschild metrics with arbitrary asymptotic behaviour.
\end{abstract}



\maketitle

\section{Introduction}

All of us have grown up with the notion that the metric outside a spherical distribution of matter is provided by the Schwarzschild solution, as proved by Birkhoff theorem \cite{Birkhoff}. Only if the object possesses an electric charge, the exterior metric is actually described by the Reissner Nordstr\"om solution.

Inside the matter distribution, the general equations for the metric and the sources were written down by Oppenheimer and Volkov \cite{Stellar} and have been used across almost one century to study the density, pressure profiles and the stability of high density stars for which general relativistic effects cannot be neglected. The interior solution and the exterior Schwarzschild solution can then be matched by standard Israel junction conditions.

In order to solve these equations, one typically assumes an equation of state $p(\rho)$ relating the pressure to the energy density of the matter under consideration. A different approach was envisaged by Tolman \cite{Tolman}, who found seven new spherically symmetric solutions by imposing conditions on the metric and then deriving the corresponding properties of the matter source. This technique to generate customized metrics has been followed and developed by many people willing to explore the possibilities offered by General Relativity without the limitations imposed by a priori assumptions on the source. Among many others, we may mention the first generating algorithm by Wyman \cite{Wyman}. Various solutions were then proposed by different authors and collected in the classification by Delgaty and Lake \cite{DelLak}. Algorithms for finding new spherically symmetric solutions were designed in different coordinates and employed in many specific problems \cite{Algorithm}.

Most of these studies focus on some minimal physical requirements for the spherically symmetric solutions proposed \cite{DelLak}. For example, they demand regularity in the origin, positive energy density, the existence of a surface boundary, defined as the value of the radial coordinate at which the pressure drops to zero. The latter condition defines isolated objects that can be matched to an exterior Schwarzschild solution. Of course, if one restricts to objects fulfilling this condition, the gravitational field at large distances is by construction identical to a standard Schwarzschild metric and there is no way to distinguish the type of matter that makes up the object. However, in the history of spherically symmetric metrics, there are interesting exceptions. Ellis illustrated the first possibility to construct a traversable wormhole within General Relativity \cite{Ellis}. This is the best known member of the class of Morris-Thorne wormholes \cite{MorTho}. Ellis wormhole does not asymptotize to Schwarzschild metric, which means that the deviations from Minkowski space do not fall as $1/r$ far from the object but rather decay as $1/r^2$. This unconventional asymptotic behaviour has immediate consequences on observable gravitational effects such as light deflection. In fact, it has been shown that the deflection angle falls down with the square of the impact parameter $b^{-2}$ \cite{EllLen,Abe,EllLen2,TakAsa}, deviating from the classical Einstein's law $\hat\alpha=4GM/c^2b$. Traversable wormholes are of great theoretical interest, since they demonstrate some potentialities of General Relativity with the inclusion of exotic forms of matter violating the energy conditions. Indeed, these violations may be allowed at quantum level in several quantum gravity theories. For these reasons, a wide literature discussing their possible observational signatures has flourished. Interestingly, with recent large scale surveys, upper limits on the number density of such exotic structures in our Universe have been set \cite{TakAsa}.

The Ellis wormhole has provided the first concrete example of a non-Schwarzschild object in General Relativity. From the point of view of gravitational lensing, this object has the surprising property of being able to defocus sources that are sufficiently aligned behind it \cite{Abe}, which is something that cannot be obtained with ordinary matter. This fact motivated the study of more general metrics that asymptotically fall as $1/r^q$, where the defocusing property has been found to be generic for $q>1$ \cite{Kit2013}. In this paper, the authors introduce a general metric falling as $1/r^q$ and exactly derive the light deflection angle, which we will recall in Sec. \ref{Section Observables}. Subsequent papers develop the gravitational lensing theory for such objects, including radial and tangential magnification \cite{Izu2014}, microlensing and centroid motion \cite{Kit2014}, strong deflection analysis \cite{Tsu2014}, and time delay \cite{Nak2014}. Signed magnifications sums were considered in Ref. \cite{TsuHar}.

With analogous phenomenological motivations coming from the chase for dark matter explanation, Ref. \cite{GalMorPec} proposed spherical solutions supported by a pure radial pressure source, without energy density. In this way, the metric assumes a logarithmic profile which can be useful to study rotation curves.

Another context in which it is difficult to match the interior solution to an exterior Schwarzschild metric is brane-world gravity \cite{MarKoy}. The regularity of solutions from the 4-dimensional brane to the 5-dimensional bulk implies that the presence of dark radiation leaking from bulk to brane is unavoidable \cite{ShMaSa,VisWil}. As a consequence, corrections to Schwarzschild are expected outside the star. In general, the metric of a Schwarzschild solution in $d$ large spatial dimensions falls as $1/r^{d-2}$ \cite{Tanghe}, also studied in Ref. \cite{Tsu2014}.

Coming back to our Universe with 3 large spatial dimensions, the existence of non-Schwarzschild objects must be supported by non-trivial sources. Ref. \cite{Kit2013} already noted that a generic metric falling as $1/r^q$ with $q>1$ implies a negative gravitational lensing convergence, i.e. an effective negative surface-mass density from the lensing point of view. The primary purpose of the present paper is to investigate the profound implications of a non-Schwarzschild metric from the point of view of the energy-momentum tensor. Understanding the (un)physical requirements needed to support such metrics is important to establish possible links from these spherically symmetric spaces introduced from a completely phenomenological perspective and fundamental physics on the other side. The physical nature of the objects described by general spherically symmetric metrics depends on the properties of the energy-momentum tensor. These can be occasionally traced to perfect fluids or scalar fields, and thus the solutions may serve as models for boson or fermion stars \cite{Boson}, sometimes advocated to explain dark matter concentrations. Particularly interesting are those solutions that are implemented by scalar or magnetic fields with suitable self-interaction potentials, since they demonstrate how these unconventional properties may indeed be obtained by fundamental fields \cite{Bron}. The existence of all these solutions, rooted in high-energy extensions of the standard model or motivated by quantum gravity, encourage our investigation of non-Schwarzschild objects from the energy conditions to the possible astrophysical implications. In fact, in the leap from purely theoretical speculations to possible astrophysical measurements, e.g. by gravitational lensing, of alternatives to Schwarzschild, the weak field limit is of the greatest importance. If the weak field limit of our metric coincides with Schwarzschild, then probe masses will follow Keplerian rotation curves and the deflection angle will invariably follow Einstein's deflection law, save for extremely small and subdominant corrections \cite{DamEsp,WDL}. Sizeable deviations from Einstein's law are only possible for metrics that do not match Schwarzschild at radial infinity, as is the case for Ellis wormhole.

Taking the move from this fact, in this work we study the interesting problem of non-Schwarzschild objects in full generality scrutinizing the implications on the source nature on one side and discussing observational signatures that may be searched for in astrophysical surveys on the other side, e.g along the lines of Ref. \cite{TakAsa}. In Section II we introduce the general asymptotic form for the metric of a spherically symmetric object and establish the links to the components of the energy-momentum tensor. In Section III we analyze the energy conditions for the sources of such metrics. In Section IV we derive the gravitational force, the redshift, and revisit the gravitational deflection. We identify some interesting families of solutions with characteristic properties in Section V and try to find relations with known exact solutions. In particular, in Section VI we explicitly show how a minimally coupled scalar field with a simple power-law potential can support solutions with arbitrary asymptotic behaviour. Section VII contains a final discussion and the conclusions of our study.

\section{Asymptotically flat metrics} \label{Sec Asy}

Let us start from a spherically symmetric metric in the form
\begin{equation}
ds^2=A(r)c^2dt^2-B(r)dr^2-r^2C(r)(d\theta^2 + \sin^2\theta d\phi^2). \label{metric}
\end{equation}

Since we are interested in the weak field limit only, for all functions $A(r),B(r),C(r)$ we just keep the slowest decaying term, that we assume to take the form of a power-law \cite{Kit2013}
\begin{eqnarray}
A(r)=1-\frac{\alpha}{r^q}+o\left(\frac{1}{r^q}\right) && \label{A(r)}\\ B(r)=1+\frac{\gamma}{r^q}+o\left(\frac{1}{r^q}\right)&& \label{B(r)}\\
C(r)=1+\frac{\beta}{r^q}+o\left(\frac{1}{r^q}\right).&&\label{C(r)}
\end{eqnarray}

Schwarzschild metric is recovered by $q=1,\alpha=\gamma=2GM/c^2,\beta=0$. Ellis wormhole is obtained setting $q=2,\alpha=\gamma=0,\beta=a^2$. We do not consider any additional effects, such as charge in Reissner-N\"ordstrom black holes, since they would just produce higher order corrections to the asymptotic metric and then to the observables \cite{WDL}.

Note that we still have the freedom to do a transformation for the radial coordinate that leaves the metric in the same form
\begin{equation}
\tilde r = r\left(1 + \frac{\xi}{r^q} \right).
\end{equation}

The coefficients of the new metric after the transformation are
\begin{equation}
\tilde \alpha = \alpha, \; \; \tilde \beta = \beta -2\xi, \; \; \tilde \gamma = \gamma + 2\xi(q-1).
\end{equation}

Then we can get rid of $C(r)$ by choosing $\xi=\beta/2$ and go to Schwarzschild coordinates with $\tilde \gamma=\gamma+(q-1) \beta$. In alternative, we can choose isotropic coordinates $B(r)=C(r)$ by letting $\xi=(\beta-\gamma)/2q$ and obtain $\tilde \gamma=\tilde\beta = \left[\gamma+(q-1) \beta \right]/q$. Indeed, in the following analysis we will adopt this prescription so as to simplify the equations and set $\beta=\gamma$.

Plugging the metric (\ref{metric}) in Einstein equations and saving only the highest power for $r$ that dominates at infinity, we obtain
\begin{eqnarray}
&& G_t^t=\frac{q(1-q)\gamma}{r^{q+2}}=\kappa \rho \label{Gt} \\
&& G_r^r=\frac{q(\gamma-\alpha)}{r^{q+2}}=-\kappa p_r \label{Gr}\\
&& G_\theta^\theta=G_\phi^\phi=\frac{q^2(\alpha-\gamma)}{2r^{q+2}}=-\kappa p_t, \label{Gth}
\end{eqnarray}
with all remaining equations being identically zero. Here we have set $\kappa=8\pi G/c^4$. The non-zero components of the energy-momentum tensor are the energy density $\rho=T_t^t$, the radial pressure $p_r=-T_r^r$, and the tangential pressure $p_t=-T_\theta^\theta=-T_\phi^\phi$.

Indeed, for $q=0$, these equations are satisfied with $\rho=p_r=p_t=0$, i.e we recover empty Minkowski space. Schwarzschild metric is recovered for $\alpha=\gamma,q=1$, which corresponds to the only other possible vacuum solution $\rho=p_r=p_t=0$.

Now let us abandon known lands and start the exploration of arbitrary power-law solutions inhabited by more or less exotic forms of matter.

A first fundamental result is that in order to support a generalized power-law weak field limit, all components of the energy-momentum tensor must be in linear relation each other. In fact, the ratios of Eqs. (\ref{Gt})-(\ref{Gth}) are all independent of the radial coordinate $r$ and can be used to derive equations of state in the form $p_r=w_r \rho,p_t=w_t\rho$, with $w_r$ and $w_t$ given by
\begin{equation}
w_r=\frac{\gamma-\alpha}{\gamma(q-1)}, \; \; w_t=\frac{q(\alpha-\gamma)}{2\gamma(q-1)}  \label{w}
\end{equation}

An interesting corollary is that the exponent $q$ determining the long-range decay of the solution is fully determined by the ratio of the tangential and radial pressures
\begin{equation}
\frac{w_t}{w_r}=-\frac{q}{2}. \label{wrwt}
\end{equation}
The only exception is the dust distribution $w_r=w_t=0$ (Section \ref{SecDust}).

\begin{figure}[h]
\includegraphics[width=7cm]{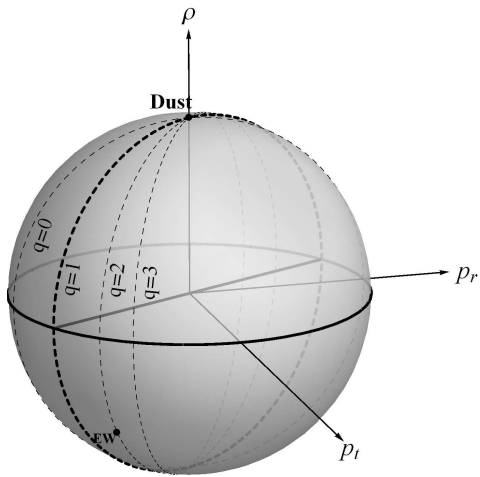}   
\caption{Graphical representation of the parameter space of generic metrics depending on the asymptotic ratios of energy density, radial and tangential pressures. Dashed circles corresponding to different asymptotic behaviours $r^{-q}$ of the metric are shown. Dust ($p_r=p_t=0$) is on top of the sphere, Ellis Wormhole ($p_r=-p_t=\rho$) is shown as EW. As noted in the text, the circle $q=1$ actually collapses to a line on the plane ($p_r$,$p_t$,$\rho=0$)} \label{Fig Sphereq}
\end{figure}

Therefore, we have a three-parameters family of weak field solutions depending on the exponent $q$ and the coefficients $\alpha,\gamma$. Alternatively, one can specify the energy density $\rho$ at some value of the radial coordinate and the two equations of state parameters $w_r,w_t$. Following the latter parametrization, we can make a graphical representation of the whole parameter space by projecting the values of $\rho$, $p_r$ and $p_t$ at a given radial coordinate to a unit sphere. In fact, the characteristics of the solutions are completely determined by the relative ratios among these quantities. Fig. \ref{Fig Sphereq} shows this unit sphere with the dashed circles corresponding to different asymptotic behaviours $r^{-q}$. In this representation, the only degeneracy arises in the case $q=1$, which necessarily implies $\rho=0$ by Eq. (\ref{Gt}). So, the circle at $q=1$ collapses to a line with $\rho=0$ on the plane ($p_r$,$p_t$). Note that the Ellis wormhole (EW in the figure) lies on its circle at $q=2$, as anticipated. The Schwarzschild solution has $\rho=p_r=p_t=0$ and would lie at the center of the sphere. Finally, by Eq. (\ref{wrwt}), we  note that the quadrants in which both pressures have the same sign ($p_r,p_t>0$ and $p_r,p_t<0$) have $q<0$, i.e. they cannot give rise to asymptotically flat metrics. This poses a severe constraint to the nature of sources that may support non-Schwarzschild metrics, and introduces us to the next section, which focuses on the energy conditions.

\section{Energy conditions}

In our asymptotic solution, the positivity of the energy density is expressed by the condition
\begin{equation}
q(1-q)\gamma\geq 0. \label{WE rho}
\end{equation}

The sign of the spatial curvature $\gamma$ is usually positive in typical spherically symmetric metrics so that the function $B(r)$ approaches 1 from above. Interestingly, a change of sign in this coefficient is not directly connected to any long-distance observables: a repulsive character of gravity may emerge only if $\alpha$ is negative, since the Newtonian potential is related to the $g_{00}$ component. Therefore, here we discuss both possibilities for the sign of $\gamma$.

In ordinary situations, with positive energy density and $\gamma>0$, the exponent $q$ is bound within the range $(0,1)$, which is what we would obtain with an extended distribution of ordinary matter (see Section \ref{SecDust}). With $q>1$, we must have either $\rho<0,\gamma>0$ or $\rho>0,\gamma<0$, i.e. something exotic must happen. An example of this situation is the Ellis wormhole, which can be written in the form (\ref{metric}), with $q=2,\alpha=\gamma=0,\beta=a^2$. Transforming to isotropic coordinates, we have  $q=2,\alpha=0,\gamma=a^2/2$. From Eqs. (\ref{Gt}), (\ref{w}), we see that indeed its source has negative energy density and equations of state $p_r=\rho,p_t=-\rho$ (see Fig. \ref{Fig Sphereq}).

The condition $\rho>0$ alone still allows for arbitrary values of the exponent $q$, provided the sign of $\gamma$ is consequently adjusted. If we further impose the weak energy condition, i.e. that for {\it any} observer on a timelike worldline the energy density is always positive, then we must have $\rho+p_r\geq 0$ and $\rho+p_t\geq 0$ everywhere. These two conditions read
\begin{eqnarray}
&& q(\alpha-q\gamma)\geq 0 \label{WE pr}\\
&& q[2\gamma-q(\alpha+\gamma)]\geq 0, \label{WE pt}
\end{eqnarray}
which complement Eq. (\ref{WE rho}).

Since we are just considering asymptotically flat metrics, $q$ must be positive and can be dropped from all inequalities.

Using Eq. (\ref{WE pr}) in Eq. (\ref{WE pt}) to eliminate $\alpha$, the last condition becomes
\begin{equation}
q(1-q)(q+2)\gamma\geq 0
\end{equation}
which is contained in (\ref{WE rho}) since $q+2>0$. Then, from Eq. (\ref{WE pr}) we learn that the weak energy condition is satisfied if and only if $\alpha\geq q\gamma$ and the energy density is positive. Once more, we note that the Ellis wormhole obviously violates this condition as well. However, with some surprise, we realize that these energy conditions still leave a lot of allowed space in the possible spherically symmetric asymptotic limits of General Relativity. We can have positive energy density and a ``wrong'' sign for $\gamma$ and still keep the weak energy condition at bay with a large enough $\alpha$.

Nevertheless, following global theorems of Ref. \cite{SchoenYau}, if the weak energy condition held anywhere, the Arnovitt-Deser-Misner (ADM) mass would be positive and the asymptotic metric would be Schwarzschild. So, even if the outer asymptotic region is supported by a source satisfying the energy conditions, in order to have $q>1$ the inner regions must necessarily violate them.

Finally, the strong energy condition stipulates that for any future pointing timelike vectors $X^\mu$ the trace of the tidal tensor $(T_{\mu\nu}-g_{\mu\nu}T/2)X^\mu X^\nu$ should be non-negative. The strongest constraint comes from $X^\mu=(1,0,0,0)$, which, for our energy-momentum tensor, gives
\begin{equation}
f\equiv (T_{00}-g_{00}T/2)=\frac{1}{2}\left(\rho+p_r+2p_t \right)\geq 0.
\end{equation}

Using Einstein equations (\ref{Gt})-(\ref{Gth}) we get
\begin{equation}
f=\frac{1}{4}q(1-q)(\alpha+\gamma) \geq 0. \label{StrongEnergy}
\end{equation}

We deduce that when the metric falls down faster than Schwarzschild ($q>1$), only if $\alpha+\gamma<0$ we can save the strong energy condition. This inequality would correspond to repulsive gravitational lensing (see Sec. \ref{Sec lensing}). By the way, we remind that any scalar field in a false vacuum actually violates the strong energy condition. So the implications of a violation at this level are not at all problematic.

\section{Observables} \label{Section Observables}

Now let us assume we have an astronomical object whose metric has the asymptotic form given by Eqs. (\ref{metric})-(\ref{C(r)}). We still stick to isotropic coordinates $\beta=\gamma$ for the description of the expected phenomenology.

\subsection{Gravitational force and rotation curves}

The gravitational force exerted by our object on a probe mass can be obtained by the geodesics equation. In the limit of slowly moving body, the normalized velocity is $\dot x^\mu\equiv (g_{00}^{-1/2},0,0,0)$ and the gravitational force is just
\begin{equation}
-{\Gamma^1}_{00}g_{00}^{-1}=-\frac{q\alpha}{2r^{q+1}}.
\end{equation}

The force decays as $r^{-q-1}$. It is attractive if $\alpha>0$ and repulsive otherwise. Central force fields with arbitrary power-laws have been studied since Newton's times. It is worth mentioning that orbits can only be closed for force fields going as $r^{-2}$ and $r$, as proved by Bertrand \cite{Bertrand}.

\begin{figure}[h]
\includegraphics[width=7cm]{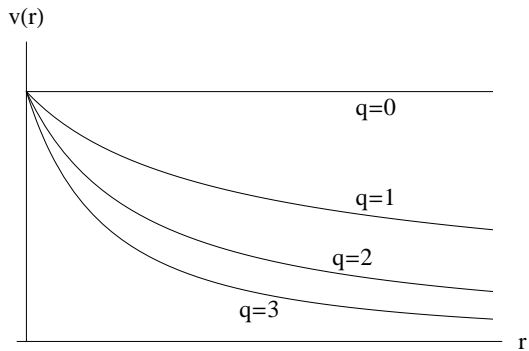}   
\caption{Rotation curves for objects rotating around an object with metric falling as $r^{-q}$. $q=0$ is the flat rotation curve typical of singular isothermal sphere distributions. $q=1$ is the classical Keplerian rotation, with $v\sim 1/\sqrt{r}$. Objects falling faster than Schwarzschild generate sub-Keplerian rotation curves.} \label{Fig RotationCurves}
\end{figure}

In this situation, to lowest order in $\alpha$, the third Kepler law for a mass orbiting at radius $r$ with orbital frequency $\omega$ is revised as
\begin{equation}
\omega^2=\frac{q\alpha}{2r^{q+2}},
\end{equation}
which implies a rotation curve
\begin{equation}
v(r)\sim r^{-q/2}.
\end{equation}
From the observational point of view, probe masses orbiting around exotic objects may follow rotation curves that are sub-Keplerian, i.e. falling faster than $1/\sqrt{r}$. In Fig. \ref{Fig RotationCurves} we draw rotation curves for several values of $q$: from $q=0$, which gives the flat rotation curve well-known for singular isothermal spheres, to $q=3$. Rotation curves around compact objects have been determined around supermassive black holes in the centers of galaxies through spectroscopic methods \cite{Sgr,Macchetto}. Of course, a sub-Keplerian rotation curve can also be explained by external perturbations, but it should be possible to distinguish the two cases. Indeed, this kinematic channel may be promising for future tests on the existence of non-Schwarzschild objects.

\subsection{Gravitational redshift and emission by accretion disks}

A photon emitted by a particle at rest ($\dot r=0$) at radius $r_e$ experiences a redshift given by
\begin{equation}
z=g_{00}^{-1/2}-1=\frac{\alpha}{2r_e^{q}}.
\end{equation}

The photon is redshifted for attractive gravity ($\alpha>0$) and blueshifted otherwise. Gravitational redshift plays a major role in determining the emission by accretion disks around black holes (see e.g. \cite{Cunningham,Disks,gfac}). The g-function, as it is sometimes called, would be steeper than Schwarzschild for $q>1$. It would be interesting to see how this change would affect the continuum spectra of accretion disks or the iron K-lines, which are an excellent probe of gravity in strong fields. However, if the accretion disk extends down to the innermost stable circular orbit (ISCO), it is mandatory to go beyond the weak field approximation and consider a full relativistic treatment for an accurate modelling of the emission which goes beyond the scope of this work.

\subsection{Gravitational lensing} \label{Sec lensing}

Gravitational lensing by non-Schwarzschild objects has been thoroughly investigated in Refs \cite{TakAsa,Kit2013,Izu2014,Kit2014,Tsu2014,Nak2014,TsuHar}. In this subsection we propose an alternative calculation of the deflection angle and time delay and compare with the results of the mentioned papers.

The deflection angle is the angle between the incoming and outgoing directions of a photon coming from infinity and returning to infinity after the close encounter with the lensing object. For a metric in the form (\ref{metric}), it is given by the following integral (see e.g. \cite{Weinberg,VirNam,LivRev,BozRev})
\begin{equation}
\hat\alpha=-\pi+2\int\limits_{b}^\infty \sqrt{\frac{B(r)}{r^2C(r)\left[\frac{r^2C(r)/A(r)}{b^2C(b)/A(b)}-1\right]}}dr,
\end{equation}
where $b$ is the minimum value of the radial coordinate reached by the photon in its journey.

Inserting our weak field expressions (\ref{A(r)})-(\ref{C(r)}), going to isotropic coordinates ($\beta=\gamma$) again and saving the lowest order terms, we get
\begin{eqnarray}
&\hat\alpha=&-\pi+2\int\limits_{b}^\infty \frac{dr}{r\sqrt{r^2/b^2-1}}  \nonumber \\ &&+\int\limits_{b}^\infty \frac{(\alpha+\gamma) r} {b^{2}\left(r^2/b^2-1\right)^{3/2}}\left(\frac{1}{b^q}-\frac{1}{r^q} \right)dr.
\end{eqnarray}

The first integral evaluates to $\pi$, canceling the $-\pi$ term. It just states that without deflector ($\alpha=\gamma=0$) there is no deflection.

The remaining integral gives the deflection angle in the weak field limit for a general spherically symmetric metric. The result reads
\begin{equation}
\hat\alpha=\frac{\sqrt{\pi}\Gamma[(1+q)/2]}{\Gamma[q/2]} \frac{\alpha+\gamma}{b^q}, \label{Angle}
\end{equation}
where $\Gamma[z]$ is the Euler gamma function.

This expression is perfectly compatible with that derived by Kitamura et al. \cite{Kit2013}, paying attention to the fact that they use Schwarzschild coordinates and make a conformal transformation, whereas we remain in isotropic coordinates. Moreover, we leave the integrand in the form of powers of polynomials without passing through trigonometric functions.

This expression for the deflection angle is valid for any spherically symmetric metric with an asymptotic power-law behaviour. Einstein's law for Schwarzschild metric is easily recovered setting $q=1,\alpha=\gamma=2GM/c^2$, as anticipated. The deflection by an Ellis wormhole in the weak field limit \cite{EllLen} ( $\hat\alpha=\frac{\pi}{4}\frac{a^2}{b^2}$)
is obtained for $q=2,\alpha=0,\gamma=a^2/2$. Anyhow, the exponent $q$ of the lowest order term in the expansion of the metric coefficients matches that of the deflection angle.

The Shapiro time delay experienced by the photon traveling in such metric is easily calculated in isotropic coordinates. For a photon traveling along $z$, the time elapsed for an infinitesimal path is
\begin{equation}
dt=\sqrt{\frac{g_{11}}{g_{00}}}dz/c=\left(1+\frac{\gamma+\alpha}{2r^q}\right)dz/c.
\end{equation}

The delay acquired with respect to a Minkowski metric is
\begin{equation}
\Delta t= \int\limits_{-\infty}^{+\infty} \frac{\gamma+\alpha}{2c r^q}dz = \frac{\sqrt{\pi}\Gamma[(q-1)/2]}{2\Gamma[q/2]} \frac{\alpha+\gamma}{cb^{q-1}}. \label{Shapiro}
\end{equation}

Comparing with Eq. (\ref{Angle}), we recover the identity $\hat\alpha=-c\nabla (\Delta t)$, which makes gravitational lensing compatible with Fermat's principle. The expression for the time delay given here is not valid for $q=1$, in which case the integral gives the familiar $-(\alpha+\gamma)\ln b$ of standard gravitational lensing \cite{SEF}. We finally note that the time delay of metrics with $1/r^q$ has been previously calculated by Nakajima et al. \cite{Nak2014}, although they left their result in integral form, save for integer $q$. Here we give a simpler expression and make the connection to the deflection angle clear.

Along the development of the theory of gravitational lensing of $1/r^q$ objects, Kitamura et al. \cite{Kit2013} also derived the so-called convergence, defined as
\begin{equation}
\sigma = -\frac{1}{2}\nabla^2(\Delta t)=\frac{1}{2} \nabla \cdot \hat{\vec\alpha},
\end{equation}
where $\hat{\vec\alpha} \equiv \hat \alpha \vec x/b$ is the vectorial form of the deflection taking into account its direction on the plane orthogonal to the line of sight ($ b=|\vec x|$ being the impact parameter). The operator $\nabla$ is defined as $\nabla \equiv (\partial_{x_1}, \partial_{x_2})$ and we are using the symbol $\sigma$ for the convergence rather than the more conventional $\kappa$ appearing in the literature since we have already used this letter for the gravitational coupling constant. Using Eq. (\ref{Angle}) we obtain
\begin{equation}
\sigma=\frac{\sqrt{\pi}\Gamma[(1+q)/2]}{\Gamma[q/2]} \frac{(1-q)(\alpha+\gamma)}{b^{q+1}}, \label{Convergence}
\end{equation}
which matches the result of Ref. \cite{Kit2013}.

In standard gravitational lensing, convergence is proportional to the mass density projected along the line of sight $\int \rho dz$. Then the authors of Ref. \cite{Kit2013} note that $q>1$ and attractive deflection ($\alpha+\gamma>0$) imply a negative convergence and then an effective negative mass density. This was interpreted as a further warning that exotic matter is needed.

In order to confirm this guess, we should resort to a general-covariant formalism for gravitational lensing taking into account all components of the energy-momentum tensor, along the lines of e.g. Ref. \cite{KopSch} or \cite{GalMor}. Neglecting any velocities of lenses, source and observer, following the first of these references, the deflection angle can be generally expressed as
\begin{eqnarray}
&\hat{\vec \alpha}(\vec x)&=\frac{\kappa}{2\pi}\int d^2x'\frac{\vec x-\vec x'}{|\vec x-\vec x'|^2} k^\alpha k^\beta  \nonumber \\&&\times\int dz \left(T_{\alpha\beta}(\vec x',z)-\frac{1}{2} \eta_{\alpha\beta} {T^\lambda}_\lambda(\vec x',z)\right),
\end{eqnarray}
where $k^\alpha=(1,0,0,1)$ is the null velocity vector of the unperturbed light ray that we are assuming to move along $z$ in cartesian coordinates; $\eta_{\mu\nu}$ is the background Minkowski metric.

The convergence is thus
\begin{eqnarray}
&&\sigma(\vec x)=\frac{1}{2} \nabla \cdot \hat{\vec\alpha}  \nonumber \\ && = \frac{\kappa}{2}k^\alpha k^\beta  \int dz \left(T_{\alpha\beta}(\vec x,z)-\frac{1}{2} \eta_{\alpha\beta} {T^\lambda}_\lambda(\vec x,z)\right)
\end{eqnarray}

Plugging the explicit expression of the velocity vector of the light ray, we get
\begin{eqnarray}
&\sigma &=\frac{\kappa}{2}\int dz \left(T^t_t-T^z_z\right) \nonumber \\
&& = \frac{\kappa}{2}\int dz \left[\rho+ p_t+ (p_r-p_t)\frac{z^2}{b^2+z^2} \right], \label{NewConvergence}
\end{eqnarray}
where we have expressed $T^z_z\equiv - p_z$ back in our spherical coordinates and used our ansatz for the energy-momentum tensor (see Sec. 2).

Using Eqs. (\ref{Gt})-(\ref{Gth}) we can eliminate the energy-momentum components in favour of the metric functions and we can easily check that the result of the integration is once more given by Eq. (\ref{Convergence}). We note that our Eq. (\ref{NewConvergence}) coincides with Eq. (183) of Ref. \cite{GalMor}, who first obtained this general expression for the convergence in a different formalism.

From Eq. (\ref{NewConvergence}) we learn that gravitational lensing is not only sensitive to the mass, but also to the pressure of the sources. In particular, light rays feel the energy density and the longitudinal pressure encountered along their path. So, the conclusion that a negative convergence implies some violation of energy conditions is correct, since in order to have the integral in Eq. (\ref{NewConvergence}) negative, there must be at least some region with $\rho+p_z<0$ and this is a clear violation of the weak energy condition. Comparing Eq. (\ref{Convergence}) with Eq. (\ref{StrongEnergy}), we also note that a negative convergence is equivalent to a violation of the strong energy condition. So, in this class of metrics, a violation of strong energy condition implies a violation of the weak energy condition.

\section{Remarkable families}

\begin{figure}[h]
\includegraphics[width=7cm]{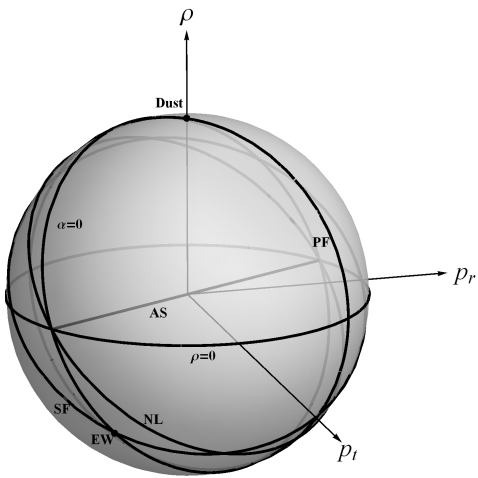}   
\caption{Families of metrics with generic asymptotic behaviour projected on the unit sphere in the space ($p_r$,$p_t$,$\rho$). {\bf EW} is the Ellis wormhole, {\bf PF} stands for perfect fluid, $\mathbf{ \rho=0}$ is the family with vanishing energy density, {\bf AS} is the pure anisotropic stress family corresponding to the degenerate $q=1$ slice, $\mathbf{ \alpha=0}$ is the family with vanishing Newtonian potential, {\bf NL} is the family with vanishing gravitational lensing, {\bf SF} is the slice obtained with minimally coupled scalar fields.} \label{Fig Spheres}
\end{figure}

In this section, we discuss some particular families of solutions that are characterized by some interesting properties, connected to the energy conditions or to the observable phenomenology. Each of these families can be identified with a slice in the unit sphere graphic representation introduced in Section \ref{Sec Asy}. Fig. \ref{Fig Spheres} collects all of them and can be a useful guide through the zoology of non-Schwarzschild asymptotic solutions.

\subsection{Perfect fluid families}

In order to recover some familiar matter sources, let us now look for perfect fluid solutions, i.e. with vanishing anisotropic stress ($p_r=p_t$). Using Eqs. (\ref{Gr}), (\ref{Gth}), this constraint reads
\begin{equation}
q(q+2)(\alpha-\gamma)=0. \label{Perfect}
\end{equation}

The case $q=0$ reduces to Minkowski, as discussed above. Let us examine the two remaining cases.

\subsubsection{Extended dust distributions} \label{SecDust}

The first family of perfect fluid solutions is obtained with $\alpha=\gamma$ in Eq. (\ref{Perfect}). Einstein equations reduce to
\begin{equation}
\kappa\rho=\frac{q(1-q)\gamma}{2r^{q+2}}, \; \; p=0.
\end{equation}

For $0<q<1$ and $\gamma>0$, the energy density is positive and we recover the weak field limit of a generic distribution of dust with density profile falling as $r^{-q-2}$. Rigorously speaking, a static distribution of particles would collapse to the center if the gravitational attraction is not balanced by any extra-forces (e.g. a repulsive charge). The solution found here corresponds to a stationary distribution of non-relativistic particles with random orbits, analogous to the case of the stars in the halos of galaxies, which typically relax to isothermal profiles ($q \sim 0$).

For $q>1$, we may have an exotic dust with negative energy density, yet we may also have ordinary matter and $\gamma<0$, as discussed above. However, in this family $\alpha=\gamma$ and we would end up with repulsive gravity ($\alpha<0$). One possible interpretation is that ordinary non-relativistic particles are shielding a central negative mass. However, a full completion of the solution down to small radii (large fields) would be needed for a full understanding of these limits.

\subsubsection{Homogeneous spaces}

Rigorously speaking, the case $q=-2$ is not asymptotically flat ($A,B,C\sim r^2$) and falls out of our starting hypothesis. Nevertheless, it is interesting to note that from Eqs. (\ref{Gt})-(\ref{Gth}) this gives uniform energy density all over the space:
\begin{equation}
\kappa\rho=-6\gamma, \;\; w=w_r=w_t=\frac{\alpha-\gamma}{3\gamma}.
\end{equation}
All possible real values of the equation of state parameter $w$ are allowed. We can immediately guess that a spherically symmetric solution with uniform energy density should be related to Friedmann-Robertson-Walker (FRW) metric. In order to make this relation manifest, we can make the perturbative transformation
\begin{eqnarray}
&& t=t'\left(1+\frac{\alpha}{2}r'^2\right) \\
&& r=r'\left(1+\frac{\alpha}{2}t'^2\right)\left(1-\frac{\gamma}{2}r'^2\right),
\end{eqnarray}
and save the linear terms in $\alpha$ and $\gamma$. We then get \begin{equation}
ds^2=c^2dt'^2-\left(1+\alpha t'^2-2\gamma r'^2\right)dr'^2-r'^2\left(1+\alpha t'^2\right)d\Omega^2,
\end{equation}
which, to linear order in $\alpha$ and $\gamma$, is equivalent to
\begin{equation}
ds^2=c^2dt'^2-a(t')^2\left[\frac{dr'^2}{1-Kr'^2}-r'^2d\Omega^2 \right]
\end{equation}
with $a(t')=1+\frac{\alpha}{2}t'^2, \; \; K=-2\gamma$.

Indeed, we find that our weak field limit with $q=-2$ is equivalent to the FRW metric of a spatially curved universe at the time of a bounce (since $\dot a(0)=0$). If $\gamma<0$, our universe has positive energy density and positive curvature, the opposite happens with $\gamma>0$.

In Fig. \ref{Fig Spheres} we can see the perfect fluid slice, labeled {\bf PF}, that falls in the two quadrants with $q<0$.

\subsection{Pure anisotropic stress family}

Digging into the mathematics of Einstein equations, we also discover that two non-trivial families of solutions exist with vanishing energy density throughout the space. We will present them here.

In the case $q=1$, Einstein equations reduce to
\begin{eqnarray}
&& G_t^t=0=\rho \\
&& G_r^r=\frac{\gamma-\alpha}{r^3}=-\kappa p_r\\
&& G_\theta^\theta=G_\phi^\phi=\frac{\alpha-\gamma}{2r^3}=-\kappa p_t,
\end{eqnarray}
which implies an equation of state $p_t=-p_r/2$. With this equation of state the energy-momentum tensor is traceless and we have a source with vanishing energy density and pure anisotropic stress, quantified by $\alpha-\gamma$.

The metric at large radii decays as $1/r$, as in Schwarzschild. Indeed, Schwarzschild solution is included in this family for the particular case $\alpha=\gamma$, corresponding to the pure vacuum solution $\rho=p_r=p_t=0$. However, as we have seen, the variety of solutions allowed for exotic sources is much richer than Schwarzschild.

In the representation of Fig. \ref{Fig Spheres}, this family is just the degenerate slice at $q=1$, which collapses to the line labeled {\bf AS} on the $\rho=0$ plane as anticipated.

\subsection{Zero spatial curvature family}

The second possibility for having vanishing energy density is to impose $\gamma=0$ in Eq. (\ref{Gt}), so that slices at constant time are flat. Here $q$ is left unconstrained, and the relation between the two pressures is  given by Eq. (\ref{wrwt}). We note that the limit $q=1$ of this family picks one member of the former family of solutions. So the two subfamilies correspond to different slices crossing at one line ($q=1,\gamma=0$ with $\alpha$ arbitrary). In Fig. \ref{Fig Spheres}, this family is the circle at $\mathbf{ \rho=0}$.

\subsection{Zero potential family}

The Ellis wormhole is characterized by $g_{00}=1$, which implies the absence of gravitational force in the Newtonian limit. The family defined by the equation $\alpha=0$ is characterized by equations of state that only depend on the exponent $q$, as $\gamma$ drops from Eqs. (\ref{w}). We have
\begin{equation}
w_r=\frac{1}{q-1}, \; \; w_t=-\frac{q}{2(q-1)}.
\end{equation}

$\gamma$ is left unconstrained and quantifies the strength of the gravitational field. Since $\alpha=0$, there is no gravitational force nor redshift at the leading order. The only physical effect that can be measured for this kind of objects is the gravitational deflection.

This family is shown as the circle labeled by $\mathbf{ \alpha=0}$ in Fig. \ref{Fig Spheres}, which also contains the Ellis Wormhole as a particular case.

\subsection{Zero lensing family}

Of course, we can have the opposite situation in which the contributions of $\alpha$ and $\gamma$ to the gravitational deflection cancel out. Imposing $\alpha=-\gamma$, the equations of state become
\begin{equation}
w_r=\frac{2}{q-1}, \; \; w_t=-\frac{q}{(q-1)},
\end{equation}
which are just twice the value they assume in the previous family.

With $\alpha>0$ for an attractive potential, $\gamma$ is negative and the energy density can be kept positive with $q>1$. This family is the circle labeled as {\bf NL} in Fig. \ref{Fig Spheres}.

\section{Minimally coupled scalar fields}

In the families visited up to now, we have imposed one property on the metric or the source equations of state and derived the consequences on the remaining variables using Einstein equations. In this subsection, we want to make a more physically motivated case by investigating the opportunities offered by a minimally coupled scalar field with a self-interaction potential. We therefore consider the following action
\begin{equation}
S=\int d^4 x \sqrt{-g}\left[\frac{R}{2\kappa}+\epsilon\left(\frac{1}{2} \partial_\mu \varphi \partial^\mu \varphi -V(\varphi) \right)\right],
\end{equation}
with $\epsilon=\pm 1$ allowing for a change in the sign of the scalar field lagrangian density.

For a static spherically symmetric configuration, $\varphi$ is a function of the radial coordinate $r$ and the energy-momentum tensor yields
\begin{eqnarray}
&& \rho=\epsilon\left[\frac{\varphi'^2}{2}\left(1-\frac{\gamma}{r^q}\right)+V(\varphi) \right] \\
&& p_r=\epsilon\left[- \frac{\varphi'^2}{2}\left(1-\frac{\gamma}{r^q}\right)+V(\varphi) \right]\\
&& p_t=-\rho,
\end{eqnarray}
where we have denoted $\varphi'\equiv \partial \varphi/ \partial r$.

The scalar field equation is
\begin{equation}
\varphi''+\frac{2\varphi'}{r}\left(1+\frac{q(\alpha-\gamma)}{4r^q} \right)-V'(\varphi)\left(1+\frac{\gamma}{r^q} \right)=0. \label{ScalarEq}
\end{equation}

Now let us consider a simple power-law potential
\begin{equation}
V(\varphi)=V_0 \varphi^n,
\end{equation}
and make the ansatz
\begin{equation}
\varphi(r)=\frac{\varphi_0}{r^m}, \; \; m>0.
\end{equation}

The dominant terms at large radii in the scalar field equation (\ref{ScalarEq}) are
\begin{equation}
m(m-1)\varphi_0 r^{-m-2}-n V_0 \varphi_0^{n-1}r^{m(1-n)}=0. \label{ScalarEq2}
\end{equation}

In order to have non-trivial solutions, we impose that the two terms are of the same order at large radii, thus obtaining a relation between the exponent $m$ and the exponent $n$ in the potential
\begin{equation}
m=\frac{1}{n/2-1}.
\end{equation}
With this value for $m$, Eq. (\ref{ScalarEq2}) is solved if
\begin{equation}
V_0=\frac{2(4-n)\varphi_0^{2-n}}{n(n-2)^2}.
\end{equation}
Note that $m>0$ only for $n>2$, which means that the scalar field vanishes at infinity only for fields with sufficiently stiff potentials.

Now we plug our ansatz in the energy-momentum tensor
\begin{eqnarray}
&& \rho=\epsilon \frac{8 \varphi_0^2}{n(n-2)^2r^{2n/(n-2)}} \label{rhosf}\\
&& p_r=-\epsilon \frac{4 \varphi_0^2}{n(n-2)r^{2n/(n-2)}}\\
&& p_t=-\rho. \label{ptsf}
\end{eqnarray}

Inserting this energy-momentum tensor in the Einstein equations (\ref{Gt})-(\ref{Gth}), and matching the exponents of $r$ at both sides, we get
\begin{eqnarray}
&& q=\frac{2}{n/2-1} \\
&& \gamma=\epsilon\kappa\frac{2\varphi_0^2}{n(n-6)} \\
&& \alpha=\epsilon\kappa\frac{(n-4)\varphi_0^2}{n(n-6)}.
\end{eqnarray}

The effective equation of states can be obtained from Eqs. (\ref{rhosf})-(\ref{ptsf}):
\begin{equation}
w_r=\frac{n}{2}-1, \; \; w_t=-1.
\end{equation}

Letting the exponent $n$ span all possible values, these equations of state describe the circle labelled by {\bf SF} in Fig. \ref{Fig Spheres}. However, the constraint $n>2$ removes the solutions in both quadrants with $p_rp_t>0$. The Ellis Wormhole is included in this family with $\epsilon<0$ and $n=4$ (a quartic self-interaction potential). More in general, we find the remarkable result that any asymptotic behaviour $r^{-q}$ for the weak field metric can be obtained just by a minimally coupled scalar field with a self-interaction potential of the form $V(\varphi)=V_0 \varphi^n$ with $n=4/q+2$. This explicit example obtained using a scalar field demonstrates that a deep investigation of all possible weak field metrics and their observable signatures, as done in Section \ref{Section Observables} is extremely precious for studies of astrophysical signatures of compact objects made up of fundamental fields.

Finally, it is interesting to note that the asymptotic forms of scalar field solutions may be much more general than those obtained with the simple power-law potential. For example, if we consider a massive complex scalar field with the potential (see \cite{Boson,Friedberg})
\begin{equation}
V(|\varphi|)=\frac{1}{2}\omega^2 |\varphi|^2+ V_0 |\varphi|^n,
\end{equation}
and an ansatz for the scalar field
\begin{equation}
\varphi(r,t)=e^{i\omega t} \varphi_0 r^{-m},
\end{equation}
the only change with respect to the previous potential comes in the energy-momentum tensor, where a mass term arises in the energy density only
\begin{equation}
\rho_\omega= \epsilon \frac{\omega^2\varphi_0^2}{r^{2/(n-1)}}.
\end{equation}
This term dominates the others (\ref{rhosf})-(\ref{ptsf}) at large radii, establishing an effective dust-like profile ($p_r,p_t \ll \rho$), with
\begin{equation}
q_\omega=\frac{2}{n/2-1}-2.
\end{equation}

It is not our intention here to explore all the host of alternatives to Schwarzschild that can be obtained by scalar or other fields. The purpose of this section was just to prove that there are substantial motivations for discussing and classifying non-Schwarzschild metrics in order to provide new links from fundamental physics to the astrophysical signatures of compact objects.

\section{Conclusions}

Looking at our Universe, the detection of any objects with a non-Schwarzschild asymptotic limit would open a sensational window towards new physics. Up to present time these objects only come out from speculations on the variety of possibilities offered by General Relativity, including travel through space and time like in the case of wormholes. Nevertheless, our ignorance on the 95\% of the energy density of our Universe under the form of Dark Matter and Dark Energy still leaves a huge space for the fantasy of theoretical physicists about the true nature of these entities. We should probably keep an open mind and admit that still much should be learned about the forms of matter allowed in our Universe.

In this spirit, we believe that our work of classification of the possible alternatives to Schwarzschild in the weak field limit may serve as a guide to astrophysical searches of compact objects made up of any kind of exotic matter. We have parameterized the solutions according to the exponent $q$ of the leading order term in the metric functions, and the values of the corresponding coefficients $\alpha$ and $\gamma$. We have investigated what these asymptotic limits imply in terms of the energy-momentum tensor, discussing the fulfillment of the energy conditions and the equations of state for radial and tangential pressure. We have also identified several interesting families of solutions with peculiar properties from the theoretical and observational point of view. It may be somewhat surprising to discover that even a minimally coupled scalar field with a power-law potential is able to support any asymptotic exponent from 0 to infinity. Given the insistence with which physicists propose scalar fields with any kind of potentials, it might be interesting to check the consequent astrophysical signatures one would find for compact objects made with such fields. In this framework, we have expressed a few observables in terms of the metric coefficients, including the gravitational force, the redshift, the lensing deflection angle and time delay. We have briefly argued about the implications in terms of rotation curves of probe masses and emission of accretion disks and discussed the implications of lensing convergence on the nature of the source of the gravitational field.

With the expressions of these observables at hand, we have the basic building blocks to confront our generalized power-law spherically symmetric solutions with the real Universe and impose constraints on the existence of exotic matter. Note that power-law deflection laws with $q\leq 1$ have been long used in strong lensing in order to fit the observed images and time delays of quasars and galaxies lensed by other galaxies or clusters of galaxies \cite{SchSlu}. Indeed, as we have shown before, a density profile of ordinary matter falling as $r^{-q-2}$ gives a potential falling as $r^{-q}$ and then a deflection angle falling as $b^{-q}$. For example, the singular isothermal sphere $\rho\sim r^{-2}$ gives $\hat\alpha=\mathrm{const}$.

We have proved that a deflection with $q>1$ would be the signature of a violation of the weak energy condition: either through a negative energy density or with spatial curvature $\gamma<0$. Gravitational lensing observables are only sensitive to the combination $\alpha+\gamma$. Only by additional measurements, such as rotation curves or gravitational redshift, it would be possible to disentangle the two coefficients and identify the nature of the source matter in full detail.

We conclude this work by warning that our weak field limits are not complete solutions of Einstein equations down to the center of the stars/objects. Our purpose, in fact, was not to demonstrate the existence of new phenomena predicted by General Relativity in strong fields, but the investigation of all possible phenomenologies allowed in the weak field limit. Prolonging these asymptotic power-law solutions to $r \rightarrow 0$ is beyond the scope of this work. Yet, we can imagine that it should not be difficult (at least numerically) to satisfy this curiosity. On the other hand, it is always fascinating to ascertain that our gravitation theory is able to subject the hypotheses of matter with unconventional properties or new physics to precise astronomically observable tests.

\begin{acknowledgments}
We thank Naoki Tsukamoto for bringing recent papers on gravitational lensing by objects falling as $1/r^{q}$ to our attention. We thank Hideki Asada for suggesting us to look into the gravitational lensing convergence. We also thank Emanuel Gallo and Osvaldo Moreschi for pointing out their works on general-covariant gravitational lensing. We finally thank Stanley Deser for pointing at the importance of global results on the ADM mass.
\end{acknowledgments}


\begin{thebibliography}{02}

\bibitem{Birkhoff} G.D. Birkhoff, "Relativity and Modern Physics", Harvard University Press, (1923).

\bibitem{Stellar} J.R. Oppenheimer and G.B. Volkov, Phys. Rev. 55, 374 (1939).

\bibitem{Tolman} R. Tolman, Phys. Rev. 55, 364 (1939).

\bibitem{Wyman} M. Wyman, Phys. Rev. 75, 1930 (1949).

\bibitem{DelLak} M.S.R. Delgaty and K. Lake, Comput. Phys. Commun. 115, 395 (1998).

\bibitem{Algorithm} S. Rahman and M. Visser, Class. Quantum Grav. 19, 935 (2002); K. Lake, Phys. Rev. D 67, 104015 (2003); D. Martin and M. Visser, Phys. Rev. D 69, 4028 (2004); P. Boonserm, M. Visser, and S. Weinfurtner, Phys. Rev. D 71, 124037 (2005); K. Lake, Phys. Rev. D 80, 4039 (2008); Th.E. Kiess, Class. Quantum Grav. 26, 5011 (2009).

\bibitem{Ellis} H. G. Ellis, J. Math. Phys. 14, 104 (1973).

\bibitem{MorTho} M. S. Morris and K. S. Thorne, Am. J. Phys. 56, 395 (1988); M. S. Morris, K. S. Thorne, and U. Yurtsever, Phys. Rev. Lett. 61, 1446 (1988).

\bibitem{EllLen} L. Chetouani and G. Cl ´ement, Gen. Relativ. Gravit. 16, 111 (1984); V. Perlick, Phys. Rev. D 69 , 064017 (2004); K. K. Nandi, Y. Z. Zhang and A. V. Zakharov, Phys. Rev. D 74, 024020 (2006); T. K. Dey and S. Sen, Mod. Phys. Lett. A, 23, 953 (2008).

\bibitem{Abe} F. Abe, ApJ, 725, 787 (2010).

\bibitem{EllLen2} A. Bhattacharya and A. A. Potapov, Modern Physics Letters A, 25, 2399 (2010); Y. Toki, T. Kitamura, H. Asada and F. Abe, Astrophys. J. 740, 121 (2011); J. M. Tejeiro and E. A. Larranaga, Rom. J. Phys. 57, 736 (2012); N. Tsukamoto, T. Harada and K. Yajima, Phys. Rev. D 86, 104062 (2012); K. Nakajima and H. Asada, Phys. Rev. D 85, 107501 (2012); G. W. Gibbons and M. Vyska, Class. Quant. Grav. 29, 065016 (2012); Ch.-M. Yoo, T. Harada, and N. Tsukamoto, Phys. Rev. D 87, 084045 (2013).

\bibitem{TakAsa} R. Takahashi and H. Asada, ApJ 768, L16 (2013).

\bibitem{Kit2013} T. Kitamura, K. Nakajima, and H. Asada, Phys. Rev. D 87, 027501 (2013).

\bibitem{Izu2014} K. Izumi, et al., Phys. Rev. D 88, 024049 (2013).

\bibitem{Kit2014} T. Kitamura, et al., Phys. Rev. D 89, 084020 (2014).

\bibitem{Tsu2014} N. Tsukamoto, T. Kitamura, K. Nakajima, and H. Asada, Phys.Rev. D 90, 064043 (2014).

\bibitem{Nak2014} 	K. Nakajima, K. Izumi, and H. Asada, Phys. Rev. D 90, 084026 (2014).

\bibitem{TsuHar} N. Tsukamoto and T. Harada, Phys. Rev. D, 87, 024024 (2013).


\bibitem{MarKoy} R. Maartens and K. Koyama, Liv. Rev. Gen. Rel. 13, 5 (2010).

\bibitem{ShMaSa} T. Shiromizu, K. Maeda and M. Sasaki, Phys. Rev. D 62, 024012 (2000).

\bibitem{VisWil} M. Visser and D.L. Wiltshire, Phys. Rev. D, 67, 104004 (2003).

\bibitem{Tanghe} F. R. Tangherlini, Nuovo Cim. 27, 636 (1963).

\bibitem{GalMorPec} E. Gallo and O.M. Moreschi, Mod. Phys. Lett. A 27, 1250044 (2012).

\bibitem{KopSch} S.M. Kopeikin and G. Sch\"afer, Phys. Rev. D 60, 124002 (1999).

\bibitem{GalMor} E. Gallo and O.M. Moreschi, Phys. Rev. D 83, 083007 (2011).

\bibitem{Boson} R. Ruffini and S. Bonazzola, Phys. Rev.
187, 1767 (1969); M. Colpi, S.L. Shapiro and I. Wasserman, Phys. Rev. D. 57, 2485 (1986); D.F. Torres, Phys. Rev. D 56, 3478 (1997); D. Spolyar, K. Freese and P. Gondolo, Phys. Rev. Lett. 100, 051101 (2008).

\bibitem{Bron} K.A. Bronnikov, Acta Phys. Pol. B4, 251, 1973; K.A. Bronnikov and  J.C. Fabris, Phys. Rev. Lett. 96, 251101 (2006); S.V. Bolokhov, K.A. Bronnikov, and M.V. Skvortsova, Class. Quantum Grav. 29, 245006 (2012).

\bibitem{DamEsp} Th. Damour and G. Esposito-Far\`ese, Phys. Rev. D, 53, 5541 (1996); J. Ebina, T. Osuga, H. Asada, and M. Kasai, Prog. of Th. Phys., 104, 1317 (2000); G.F. Lewis, X.R. Wang, Prog. of Th. Phys., 105, 893 (2001).

\bibitem{WDL} M. Sereno, Phys. Rev. D, 69, 023002 (2004); C. R. Keeton and A. O. Petters, Phys. Rev. D 72, 104006 (2005); 73, 044024 (2006).

\bibitem{SchoenYau} R. Schoen and Sh.-T. Yau, Phys. Rev. Lett. 43, 1457 (1979).

\bibitem{Bertrand}  J. Bertrand, C. R. Acad. Sci. 77, 849 (1873).

\bibitem{Sgr} A. Eckart and R. Genzel, MNRAS 284, 576 (1997); S. Gillessen et al., ApJ, 692, 1075 (2009).

\bibitem{Macchetto} F. Macchetto et al., ApJ 489, 57 (1997).

\bibitem{Cunningham} C.T. Cunningham, ApJ 202, 788 (1975).

\bibitem{Disks} A.C. Fabian, M.J. Rees, L. Stella, and N.E. White, MNRAS 238, 729 (1989); Tanaka et al., Nature 375, 659 (1995).

\bibitem{gfac} M. Dovciak M., V. Karas, and T. Yaqoob, ApJS, 153, 205 (2004); K. Beckwith and C. Done, MNRAS, 352, 353 (2004).

\bibitem{Weinberg} S. Weinberg, "Gravitation and Cosmology: Principles and Applications of the General Theory of Relativity", John Wiley \& Sons (1972).

\bibitem{VirNam} K.S. Virbhadra, D. Narasimha, and S.M. Chitre, A\&A 337, 1 (1998).

\bibitem{LivRev} V. Perlick, Living Rev. Relativity 7 , 9
(2004).

\bibitem{BozRev} V. Bozza, Gen. Rel. and Grav. 42 , 2269
(2010).

\bibitem{SEF}  P. Schneider, J. Ehlers, and E.E. Falco, {\it Gravitational lenses}, Springer-Verlag, Berlin (1992).

\bibitem{Friedberg} R. Friedberg, T.D. Lee, and Y. Pang, Phys. Rev. D35, 3658 (1987).

\bibitem{SchSlu} P. Schneider and D. Sluse, A\&A, 559, 37 (2013).


\end{thebibliography}
\end{document}